\documentclass[prl,aps,twocolumn,showpacs,floatfix]{revtex4-1}
\usepackage{amsfonts}
\usepackage{amssymb}
\usepackage{hyperref}
\usepackage{graphicx}
\usepackage{dcolumn}
\usepackage{bm,amsmath,verbatim}
\usepackage{mathrsfs}
\usepackage{color}
\usepackage{lineno}

\usepackage[T1]{fontenc}
\usepackage[latin9]{inputenc}
\usepackage{booktabs}

\hypersetup{colorlinks,
	linkcolor=blue,          citecolor=blue,        filecolor=blue,      urlcolor=blue           }

\begin{document}

\title{Quadrupole Insulator without Corner States in the Energy Spectrum}
\author{Yu-Liang Tao$^{1}$}
\author{Jiong-Hao Wang$^{1}$}
\author{Yong Xu$^{1,2}$}
\email{yongxuphy@tsinghua.edu.cn}
\affiliation{$^{1}$Center for Quantum Information, IIIS, Tsinghua University, Beijing 100084, People's Republic of China}
\affiliation{$^{2}$Hefei National Laboratory, Hefei 230088, PR China}

\begin{abstract}
The quadrupole insulator is a well-known instance of higher-order topological insulators in two dimensions, 
which possesses midgap corner states in both the energy spectrum and entanglement spectrum.
Here, by constructing and exploring a model Hamiltonian under a staggered $\mathbb{Z}_2$ gauge field 
that respects momentum-glide reflection symmetries, we surprisingly find a quadrupole 
insulator that lacks zero-energy corner modes in its energy spectrum, despite possessing a nonzero quadrupole moment.
Remarkably, the existence of midgap corner modes is found in the entanglement spectrum.
Since these midgap states cannot be continuously eliminated, the quadrupole insulator cannot be 
continuously transformed into a trivial topological insulator, thereby confirming its topological nature.
We show that the breakdown of the correspondence between the energy spectrum and entanglement spectrum 
occurs due to the closure of the edge energy gap when the Hamiltonian is flattened.
Finally, we present a model that demonstrates an insulator with corner modes in the energy spectrum even 
in the absence of the quadrupole moment. In this phase, the entanglement spectrum does not display any midgap states.
The results suggest that the bulk-edge correspondence of quadrupole insulators generally manifests in the 
entanglement spectrum rather than the energy spectrum.
\end{abstract}
\maketitle

Higher-order topological insulators have experienced rapid development in recent years~\cite{Taylor2017Science,Taylor2017PRB,Fritz2012PRL,ZhangFan2013PRL,Slager2015PRB,Brouwer2017PRL,FangChen2017PRL,
	Schindler2018SA, Wang2018EL, Brouwer2019PRX,Seradjeh2019PRB,Roy2019PRB,Yang2019PRL,Hughes2020PRB, Xu2020PPR,
	Parameswaran2020PRL,AYang2020PRL,Xu2020NJP,Roy2020PPR,Wang2021PRL,Xu2023PRB}, 
serving as a generalization of conventional first-order topological insulators.
In contrast to first-order topological states which have $m=1$, these higher-order states 
support edge states of $(n-m)$ dimensions ($1< m \le n$) in an $n$-dimensional system. 
A notable example of a higher-order topological phase in two dimensions (2Ds) 
is the quadrupole insulator~\cite{Taylor2017Science,Taylor2017PRB,Huber2018Nature,Bahl2018Nature}, 
which showcases topologically protected corner states. 
These topological insulators are identified by their quantized quadrupole moment~\cite{Cho2019PRB,Hughes2019PRB}, 
which is enforced by symmetries, such as chiral symmetry~\cite{Xu2021PRB,Shen2020PRL}.

The entanglement spectrum provides an alternative means of describing the topological 
properties of a system~\cite{Haldane2008PRL}. 
It refers to the eigenvalue spectrum of the reduced density matrix 
for a system comprised of two separate subsystems. 
While the connection between the entanglement spectrum and edge energy spectrum is 
demonstrated for conventional first-order topological insulators~\cite{Ryu2006PRB,Dodriguez2009PRB,BrayAli2009PRB, 
Pollmann2010PRB, Fidkowski2010PRL,Prodan2010PRL,Turner2010PRB,
THughes2011PRB,Alexand2011PRB,Chandran2011PRB,Bernevig2013PRB}, there are exceptions.
For example, in systems with inversion symmetry, it has been observed that when the 
energy spectrum exhibits a gap without midgap states, the entanglement spectrum contains 
midgap modes~\cite{Turner2010PRB,THughes2011PRB}. 
This gives rise to a topological state as the midgap states in the entanglement 
spectrum cannot be continuously removed. In the context of higher-order topological phases, 
the entanglement spectrum between a quarter part and its complement is considered so 
that a corner boundary is provided. It has been found that quadrupole insulators consistently harbor 
midgap modes in both the energy spectrum and entanglement spectrum~\cite{Schindler2018SA,
Wang2018EL,Hughes2020PRB,Xu2020PPR,Dubinkin2020arxiv}. 
This suggests a strong correspondence between the two spectra in higher-order topological phases.

\begin{figure}[t]
	\includegraphics[width = 1\linewidth]{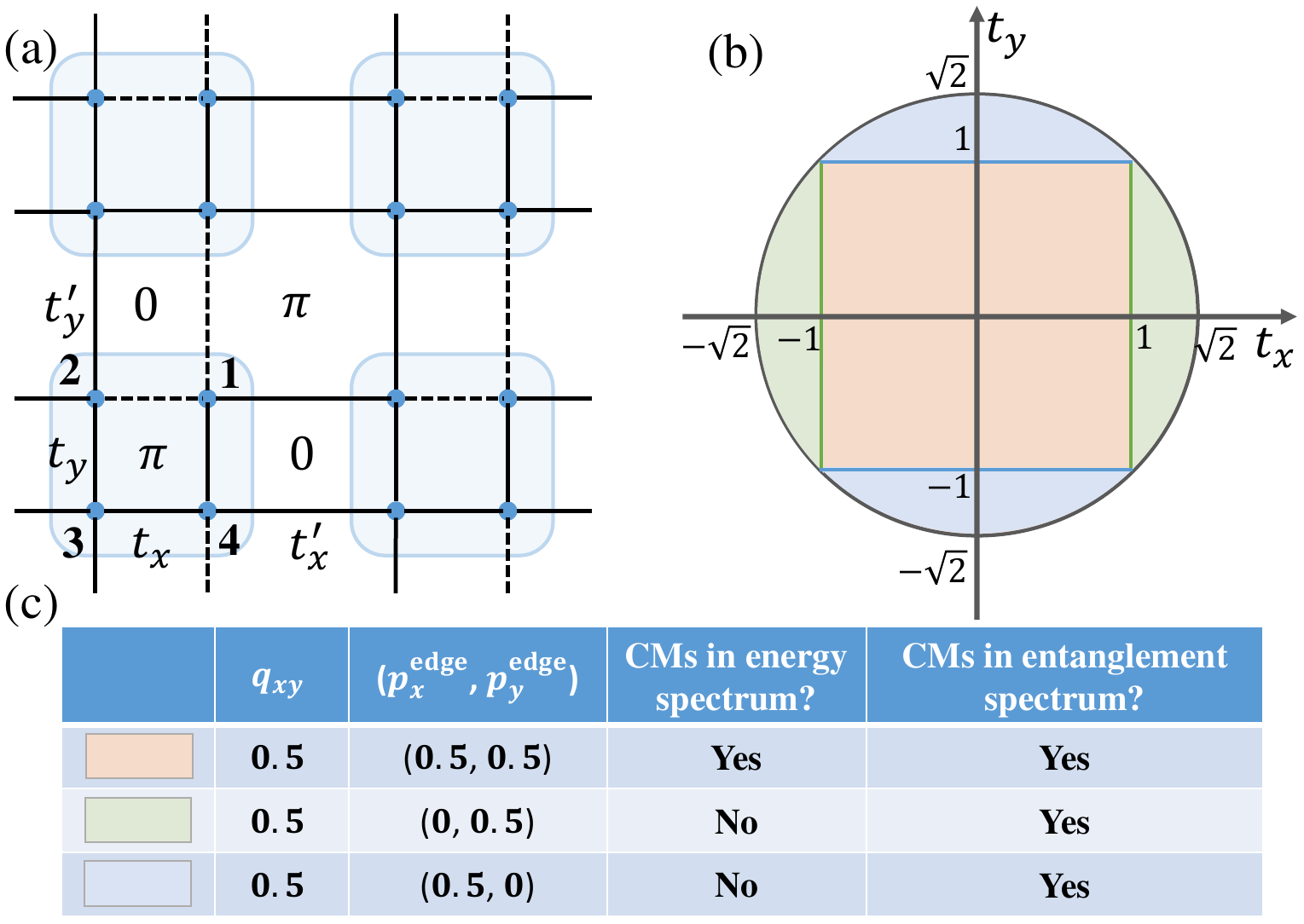}
	\caption{
		(a) Schematics of our tight-binding model with a staggered $\pi$ flux configuration respecting 
		momentum-glide reflection symmetries. In each unit cell,
		there are four sites labelled by $1,2,3,4$. The dashed lines represent the hopping with a phase of $-1$. 
		(b) Phase diagram of the model in the $(t_x,t_y)$ plane with $t_x^\prime$ and $t_y^\prime$ set to one.
		(c) Table listing the quadruple moment $q_{xy}$, the edge polarizations $(p_x^{\text{edge}},p_y^{\text{edge}})$,
		and whether corner modes (CMs) exist in the energy spectrum and entanglement spectrum 
		in each phase.
	}
	\label{fig1}
\end{figure}

In the paper, we introduce a tight-binding model under a staggered $\mathbb{Z}_2$ gauge field 
and surprisingly find an exotic topological phase. 
This phase breaks the correspondence between the energy spectrum and entanglement spectrum.
More specifically, we find that this phase possesses a nonzero quadrupole moment, indicating that it can 
be classified as a quadrupole insulator (see Fig.~\ref{fig1}). However, it does not exhibit midgap corner modes in the energy spectrum. 
Remarkably, we observe the emergence of midgap states in the entanglement spectrum. 
This result suggests that the quadrupole
moment generally identifies the presence of midgap states in the entanglement spectrum instead of
the energy spectrum.
The reason for this is that in the case of higher-order topology, the edge energy gap can close, causing corner modes 
to appear or disappear without any changes occurring in the bulk states when we flatten the Hamiltonian by 
altering the eigenenergies. Consequently, as we transition from a flattened Hamiltonian to the original one, 
the corner states vanish while the quadrupole moment remains unaffected.
As for the entanglement spectrum, since it is derived from the bulk states, it exhibits a relationship 
with the energy spectrum of the flattened version of the original Hamiltonian. To further validate our conclusion, 
we introduce a model with zero quadrupole moment and subsequently observe the presence of corner modes 
in the energy spectrum, while they are absent in the entanglement spectrum.

\emph{{\color{blue}Model Hamiltonian}}.--- 
We start by introducing a 2D tight-binding model shown in Fig.~\ref{fig1}(a) 
where there are four sites in each unit cell. The hopping is endowed with the phase of $1$ or $-1$
represented by solid or dashed lines, respectively, leading to a flux of $0$ or $\pi$ in each plaquette. 
Its Bloch Hamiltonian in momentum space is given by 
\begin{align} \label{Hk}
	H({\bm{k}})=&-t_x\tau_z\sigma_x+t_y\tau_x\sigma_x+t_x^\prime\cos{k_x}\tau_0\sigma_x \\ \nonumber
	&-t_x^\prime\sin{k_x}\tau_z\sigma_y+t_y^\prime\cos{k_y}\tau_y\sigma_y+t_y^\prime\sin{k_y}\tau_x\sigma_y,
\end{align}
where $\tau_i$ and $\sigma_i$ with $i=x,y,z$ are Pauli matrices, $\tau_0$ and $\sigma_0$ are $2\times2$ identity matrices, 
and their tensor products act on the internal degrees of freedom in the unit cell. 
$t_\nu$ and $t_\nu^\prime$ with $\nu=x,y$ denote the intracell and intercell hopping strengths along the $\nu$ direction.
For simplicity, we choose $t_x^\prime=t_y^\prime=1$ as the units of energy.  
The system respects time-reversal symmetry $T=\kappa$ ($\kappa$ is the complex conjugate operator), 
i.e., $TH(\bm{k})T^{-1}= H(-\bm{k})$, and chiral symmetry $\Gamma=\tau_0\sigma_z$, i.e., 
$\Gamma H(\bm{k})\Gamma^{-1}=-H(\bm{k})$. 
Chiral symmetry acts as a protective mechanism to ensure the quantization of the quadrupole moment
~\cite{Xu2021PRB,Shen2020PRL}, making it a robust and well-defined topological invariant.
Due to the staggered $\mathbb{Z}_2$ gauge field, 
two momentum-glide reflection symmetries~\cite{Yuxin_KB,Tao_KB,ZhuArxiv, ChengArxiv}, $M_x=\tau_0\sigma_x$ and $M_y=\tau_y\sigma_y$, 
are respected so that $M_x H(k_x,k_y)M_x^{-1}=H(-k_x,\pi+k_y)$ and $M_y H(k_x,k_y)M_y^{-1}=H(\pi+k_x,-k_y)$. 

\begin{figure}[t]
	\includegraphics[width = 1\linewidth]{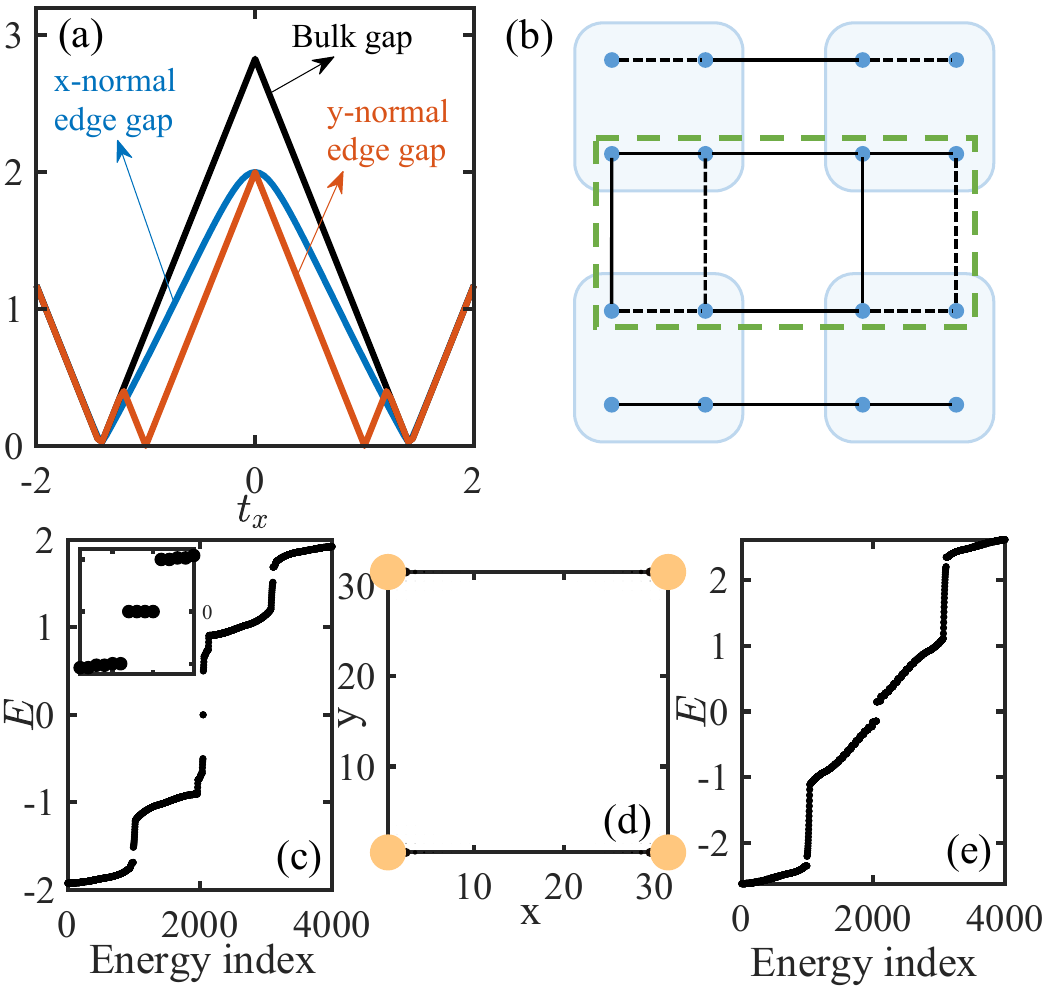}
	\caption{
		(a) The bulk energy gap (black line), $x$-normal edge energy gap (blue line) and $y$-normal edge energy gap (red line)
		versus $t_x$ when $t_y=0$. The $y$-normal edge energy gap closes at $t_x=\pm 1$ while the bulk energy gap closes at $t_x=\pm\sqrt{2}$.
		(b) Schematics of our model when $t_y=0$.
		(c) The energy spectrum of our model under open boundary conditions (OBCs) at $t_x=0.5$ and $t_y=0.1$,
		illustrating the presence of four zero-energy modes. The inset shows its zoomed-in view near zero energy.
		(d) The spatial distribution of zero-energy states identified by $n(\bm{r})=\sum_{j=1}^4 \sum_{\alpha}|\phi_j(\bm{r},\alpha)|^2$
		where $\phi_j(\bm{r},\alpha)$ is the $\alpha$th component of the wave function of the $j$th zero-energy mode 
		at unit cell $\bm{r}$.
		(e) The energy spectrum of our model under OBCs at $t_x=1.2$ and $t_y=0.1$, showing the absence of
		zero-energy modes. 
	}
	\label{fig2}
\end{figure}

Now, we utilize the quadrupole moment and edge polarizations to characterize the higher-order nontrivial topology. 
The quadrupole moment is defined by~\cite{Cho2019PRB,Hughes2019PRB,Xu2021PRB,Shen2020PRL}
\begin{align} \label{Qxy}
	q_{xy}=\left[\frac{1}{2\pi}\text{Im} \log \det (U_o^\dagger \hat{D} U_o)-Q_0\right]\ \text{mod}\ 1,
\end{align}
where $U_o=(|\psi_1\rangle,\dots,|\psi_{2L^2}\rangle)$ with $|\psi_j\rangle$ 
being the $j$th occupied eigenstate of an $L\times L$ system under periodic boundary conditions (PBCs) with $j=1,\dots,2L^2$, 
and $\hat{D}=\text{diag}\left\{e^{2\pi i x_j y_j/L^2}\right\}_{j=1}^{4L^2}$ with $(x_j,y_j)$ being the spatial position of lattice site $j$. 
Here, $Q_0$ is contributed by background positive charges. 
In order for a system to possess a well-defined quadrupole moment, it is essential to ensure the absence 
of bulk dipole moments, which is achieved by momentum-glide reflection symmetries $M_x$ and $M_y$.
When $q_{xy}=0.5$, the nontrivial phase is classified as a quadrupole insulator. 

To characterize the nontrivial edge property on an $L_x\times L_y$ lattice, 
we calculate the edge polarization $p_x^{\text{edge}}$ along $x$ 
(similarly for $p_y^{\text{edge}}$ along $y$) 
in a cylinder geometry with open boundaries along $y$.
It is determined by the sum of the distribution of polarization over a half lattice along $y$~\cite{Taylor2017Science,Taylor2017PRB,Xu2020PPR}, i.e., 
\begin{align} \label{pedge}
	p_x^{\text{edge}}=\sum_{R_y=1}^{L_y/2}p_x(R_y).
\end{align}
Here, $p_x(R_y)$ is the distribution of polarization at cell $R_y$.
We calculate $p_x(R_y)$ based on
\begin{align} \label{pd}
	p_x(R_y)=\sum_{j=1,2}\rho^j(R_y)\nu_x^j,
\end{align}
where $\rho^j(R_y)=(1/L_x)\sum_{k_x,\alpha}|\sum_{n=1}^{2L_y}[u_{k_x}^n]^{R_y,\alpha}[\nu_{k_x}^j]^n|^2$ 
is the probability distribution of the hybrid Wannier functions, 
$[\nu_{k_x}^j]^n$ is the $n$th component of the $j$th eigenstates of the Wannier Hamiltonian with eigenvalue $\nu_{x}^j$, 
and $[u_{k_x}^n]^{R_y,\alpha}$ is the component of the $n$th occupied eigenstates of the Hamiltonian $H(k_x,L_y)$. 

\emph{{\color{blue}Phase diagram}}.--- 
We map out the phase diagram with respect to $t_x$ and $t_y$ in Fig.~\ref{fig1}(b) based on the quadrupole moment
and edge polarizations. We find that when $t^2=t_x^2+t_y^2<2$, $q_{xy}=0.5$, indicating that the phase corresponds to
a quadrupole insulator, which is in stark contrast to the the Benalcazar-Bernevig-Hughes (BBH) model where 
a quadrupole insulating phase appears in the square region (light red region) with $|t_x|<1$ and $|t_y|<1$. 
Such a phase arises from a topologically trivial phase with $q_{xy}=0$ 
through a bulk energy gap closure at $t^2=2$ as we decrease $t^2$ [see the black line in Fig.~\ref{fig2}(a)]. 
However, we surprisingly find that zero-energy
corner modes in the energy spectrum only exist in the red region where $p_x^{\text{edge}}=p_y^{\text{edge}}=0.5$
as shown in Fig.~\ref{fig2}(c)--(d).
In the green and blue regions, although $q_{xy}=0.5$, no midgap corner modes appear in the energy spectrum as
illustrated in Fig.~\ref{fig2}(e). In addition, in the green region, 
$(p_x^{\text{edge}},p_y^{\text{edge}})=(0,0.5)$ and in the blue region, $(p_x^{\text{edge}},p_y^{\text{edge}})=(0.5,0)$.
While type-II quadrupole insulators exhibit the same edge polarization configurations, they also possess corner modes 
in contrast to this case. In fact, these phases also satisfy
the relation that $Q^{\mathrm{corner~}}=p_y^{\mathrm{edge~}}+p_x^{\mathrm{edge~}}-q_{xy}$
with $Q^{\mathrm{corner~}}$ being the corner charge~\cite{Taylor2017Science,Taylor2017PRB}, while the type-II one violates it~\cite{Xu2020PPR}.

\begin{figure}[t]
	\includegraphics[width = 1\linewidth]{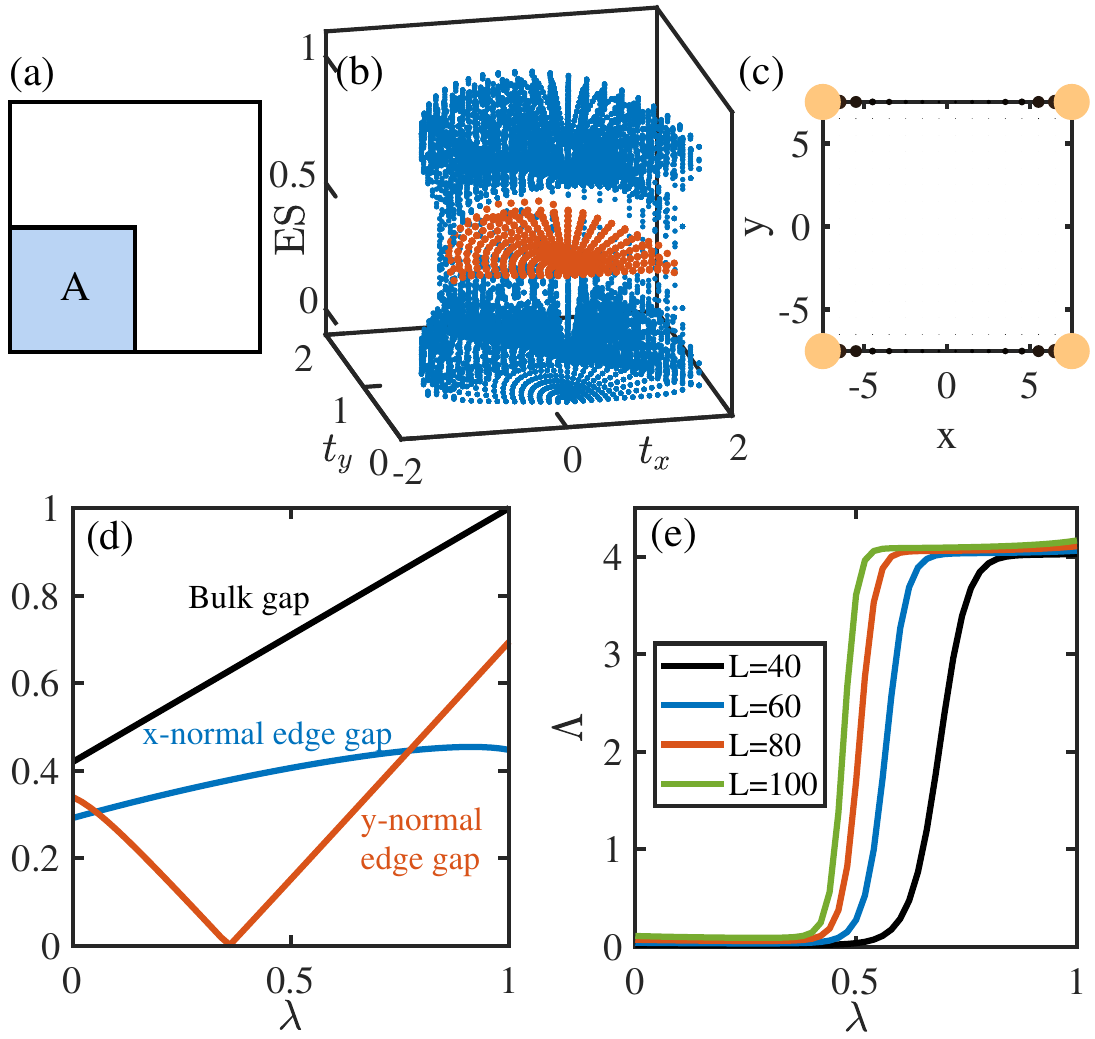}
	\caption{
		(a) Schematics showing that the entanglement spectrum between a quarter part $A$ and its complement is computed. 
		(b) The entanglement spectrum in the $(t_x,t_y)$ plane where midgap states at $\xi=0.5$ are highlighted
		as red colors. For visual clarity, we only plot the part with $t_y>0$. 
		(c) The spatial distribution of the midgap states in (b) at $t_x=1.2$ and $t_y=0.1$.
		(d) The bulk energy gap (black line), $x$-normal edge energy gap (blue line) and $y$-normal edge energy gap (red line)
		as a function of $\lambda$, showing the closure of the y-normal edge energy gap at $\lambda=0.36$. 
		(e) The relative zero-energy DOS defined as $\Lambda=\rho(0)/F_\delta$ for the model $H_{\text{def}}(\lambda)$ under OBCs,
		where $\rho(E=0)=\sum_j \delta(E-E_j)$ is the zero-energy densify of states (DOS) and $F_\delta$ is the maximum value of 
		the $\delta$ function in numerical calculations. The results for distinct system sizes suggest that zero-energy corner
		modes appear once the edge energy gap closes and reopens.
	}
	\label{fig3}
\end{figure}

These phases transition into the traditional quadrupole insulator through the closure of an edge energy gap  
[see Fig.~\ref{fig2}(a)], leading to the change of one edge polarization while preserving the quadrupole moment.
For instance, consider $t_y=0$ so that the model reduces to the one shown in Fig.~\ref{fig2}(b).
Clearly, the $y$-normal edge states are described by the Su-Schrieffer-Heeger (SSH) model which 
experiences an energy gap closing at $t_x=\pm1$. Such a gap closure results in a change in
$p_x^{\text{edge}}$ from $0$ to $0.5$.
However, the bulk energy gap vanishes when $t_x=\pm \sqrt{2}$ at $k_x=0,\pi$
as seen from the bulk energies, $E_{\textbf{b},\pm}^2=t_x^2+2 \pm 2|t_x|\sqrt{1+\cos^2 k_x}$.
This differs from the BBH model whose spectrum always remains gapped as seen from its 
energies, $E_{\textbf{b}}^2=t_x^2+2t_x\cos k_x+2$.

We now evaluate the entanglement spectrum
by diagonalizing the correlation matrix in a quarter
subsystem $A$ as shown in Fig.~\ref{fig3}(a) defined as~\cite{Peschel2003}
\begin{align} \label{corre}
	\left[C_A\right]_{\bm{r}_i\alpha,\bm{r}_j\beta}=\langle\hat{c}^\dagger_{\bm{r}_i\alpha}\hat{c}_{\bm{r}_j\beta}\rangle,
\end{align}
where $\hat{c}^\dagger_{\bm{r}_i\alpha}$ ($\hat{c}_{\bm{r}_i\alpha}$) is the fermionic creation (annihilation) operator at lattice site 
$(\bm{r}_i,\alpha)$. Figure~\ref{fig3}(b) illustrates that midgap modes exist in the entanglement 
spectrum in the region with $q_{xy}=0.5$. These modes are mainly localized at corners as shown in 
Fig.~\ref{fig3}(c). In fact, the entanglement spectrum has a one-to-one correspondence with the 
energy spectrum of the flattened Hamiltonian with open boundaries enclosing the quarter part~\cite{supplement},
similar to the first-order case~\cite{Fidkowski2010PRL,Turner2010PRB,THughes2011PRB}.
Since the quadrupole moment is evaluated using bulk states, it describes the topology of a 
flattened Hamiltonian. Our results thus indicate that the bulk-edge correspondence of quadrupole insulators 
generally manifests in the entanglement spectrum rather than the energy spectrum. 
This prompts the question of why there might be a breakdown of the bulk-edge relationship 
in the energy spectrum of the original Hamiltonian.

We find that the breakdown occurs because the edge energy gap can close even though the bulk states remain
unchanged as the Hamiltonian is flattened. Specifically, we define 
$H_{\text{def}}(\lambda)=UD(\lambda)U^\dagger$, 
where $U=(|\psi_1\rangle,\dots,|\psi_{4L^2}\rangle)$ with $|\psi_j\rangle$ being the $j$th eigenstate 
of our model under PBCs. Eigenenergies of $H_{\text{def}}(\lambda)$ are listed in
$D(\lambda)=\lambda\text{diag}(-0.5,0.5)\otimes I_{2L^2}+(1-\lambda) \text{diag}(E_1,\dots,E_{4L^2})$
with $E_1,\dots,E_{4L^2}$ being eigenenergies of our model in Eq.~(\ref{Hk}) 
sorted in an ascending order. As we vary $\lambda$ from $0$ to $1$, the Hamiltonian is continuously deformed 
from the original Hamiltonian to the flattened one without involving bulk energy gap closing [see Fig.~\ref{fig3}(d)].
Since the eigenstates remain unchanged during the process, the quadrupole moment does not change.
However, we find that an energy gap closure at $y$-normal
boundaries occurs for $H_{\text{def}}(\lambda)$ under OBCs along $y$ [see Fig.~\ref{fig3}(d)],   
rendering the emergence of corner modes as reflected by a relative zero-energy 
DOS [see Fig.~\ref{fig3}(e)].
 
\begin{figure}[t]
	\includegraphics[width = 1\linewidth]{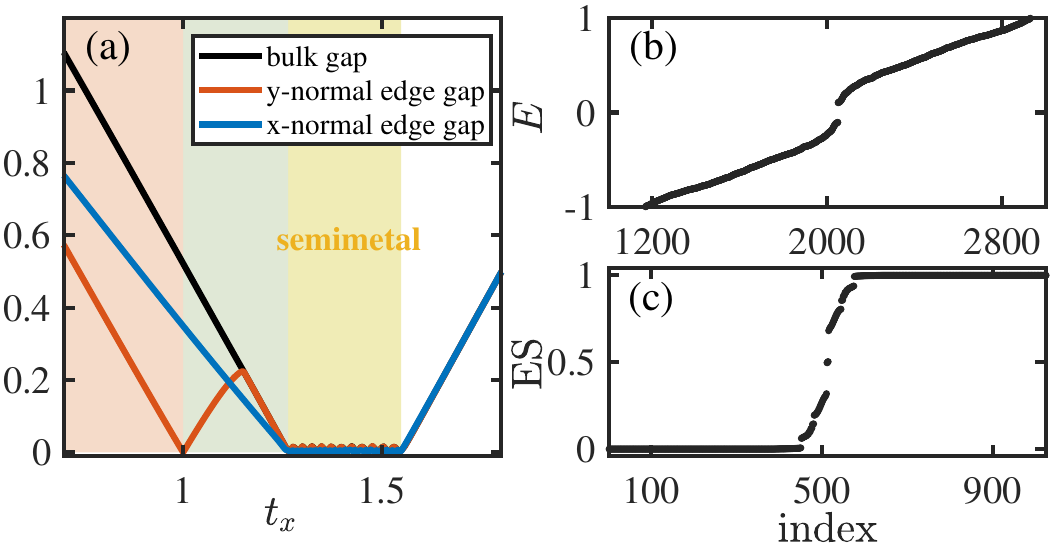}
	\caption{
		(a) The bulk energy gap (black line), $y$-normal edge energy gap (red line) and $x$-normal edge energy gap 
		(blue line) versus $t_x$ at $t_y=0$ for the Hamiltonian in Eq.~(\ref{Hk}) with two extra terms 
		$\Delta_1\tau_y\sigma_x$ and $\Delta_2\tau_y\sigma_y$ added. Here, we take $\Delta_1=\Delta_2=0.2$.
		The red and blue regions refer to the corresponding phases in Fig.~\ref{fig1}(b).
		Besides, there appears a semimetallic phase in the yellow region hosting four Dirac points
		in the momentum-space energy spectrum.
		(b) The energy spectrum and (c) the entanglement spectrum 
		for the Hamiltonian in (a) at $(t_x, t_y)=(1.1,0)$.
		In (b), OBCs are applied along both $x$ and $y$.
		The figure illustrates that despite the absence of midgap states in (b), they appear in (c).
			}
	\label{fig4}
\end{figure}

\emph{{\color{blue}Model without momentum-glide reflection symmetry}}.--- 
Although the Hamiltonian in Eq.~(\ref{Hk}) respects time-reversal symmetry and momentum-glide reflection
symmetries, they are not essential to the quadrupole insulator. To clarify this, we add 
two extra terms $\Delta_1\tau_y\sigma_x$ and $\Delta_2\tau_y\sigma_y$, breaking these
symmetries while preserving chiral symmetry. We have also checked that the bulk dipole 
moments are zero so that the quadrupole moment is well defined. 
In this case, we still observe the presence of the phases represented by the
same color as in Fig.~\ref{fig1}(b) identified by the energy gaps, 
quadrupole moment and edge polarizations [see Fig.~\ref{fig4}(a)]. 
In the green region, while $q_{xy}=0.5$, no midgap modes are observed in the energy 
spectrum [see Fig.~\ref{fig4}(b)]. However, they arise in the entanglement spectrum 
shown in Fig.~\ref{fig4}(c).
Interestingly, besides these phases, we also observe a semimetal phase (yellow region) with 
four Dirac points in momentum-space energy spectra.

\begin{figure}[t]
	\includegraphics[width = 1\linewidth]{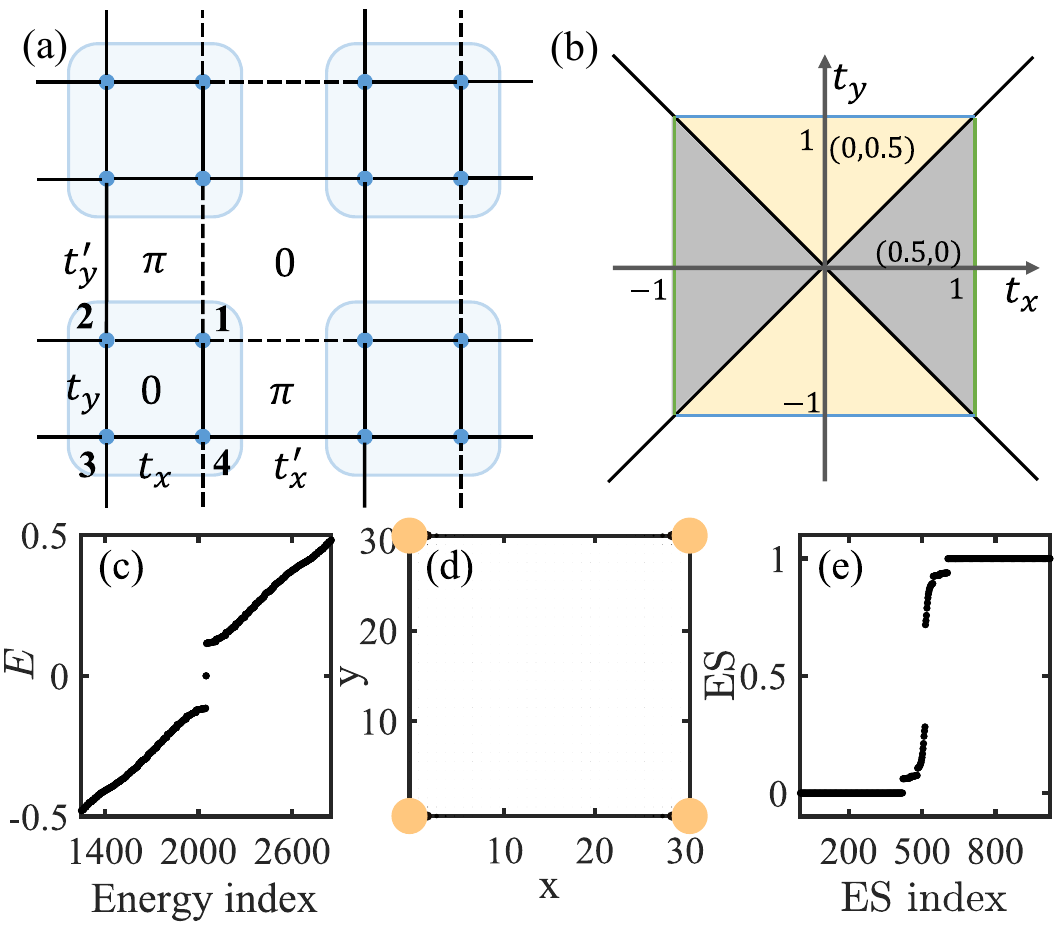}
	\caption{
	(a) Schematics of the tight-binding model in Eq.~(\ref{ano_Hk}).
	(b) Phase diagram of the model in (a). In all these phases, $q_{xy}=0$. 
	In the 
	gray (gold) region, $(p_{x}^{\text{edge} },p_{y}^{\text{edge} })=(0.5,0)$ [$(0,0.5)$], and
	in other regions, $(p_{x}^{\text{edge} },p_{y}^{\text{edge} })=(0,0)$.
	In both the gray and gold regions, midgap corner modes exist in the energy spectrum 
	[see (c) for the energy spectrum and (d) for the spatial distributions of corner modes],
	whereas they do not exist in the entanglement spectrum shown in (e). 
	In (c)-(e), $t_x=0.5$ and $t_y=0.1$. On the boundaries highlighted as green and blue
	lines, one of edge energy gaps vanishes, and on the black lines, the bulk energy gap closes.
	}
	\label{fig5}
\end{figure}

\emph{{\color{blue}Model without quadrupole moment}}.---  
Next, we construct another model as shown in Fig.~\ref{fig5}(a). Similar to the model 
in Eq.~(\ref{Hk}), this model is still subject to staggered $\mathbb{Z}_2$ gauge fields;
however, each unit cell does not carry a $\pi$ flux. The Bloch Hamiltonian in momentum space reads 
\begin{align} \label{ano_Hk}
	H_{2}({\bm{k}})=&t_x\tau_0\sigma_x+t_y\tau_x\sigma_x-t_x^\prime\cos{k_x}\tau_z\sigma_x \\ \nonumber
	&+t_x^\prime\sin{k_x}\tau_0\sigma_y+t_y^\prime\cos{k_y}\tau_y\sigma_y+t_y^\prime\sin{k_y}\tau_x\sigma_y.
\end{align}
It still respects time-reversal symmetry $T=\kappa$, chiral sysmmetry $\Gamma=\tau_0\sigma_z$, 
and two momentum-glide reflection symmetries, $M_x=\tau_0\sigma_x$ and $M_y=\tau_x\sigma_x$. 
Similarly, the two reflection symmetries enforce the absence of bulk dipole moments so that the quadrupole moment
is well defined. We also take $t_x^\prime=t_y^\prime=1$ as the units of energy.

We find that the model always has zero quadrupole moment, implying that it is a trivial quadrupole 
insulator. Consequently, the absence of midgap modes in the entanglement spectrum is observed in 
Fig.~\ref{fig5}(e), consistent with the bulk-edge correspondence of quadrupole insulators in 
the entanglement spectrum.
However, midgap corner modes appear in the energy spectrum as revealed in Fig.~\ref{fig5}(c)--(d) 
when $|t_x|<1$ and $|t_y|<1$ corresponding to the grey and gold regions in Fig.~\ref{fig5}(b). 
These two regions also exhibit the edge polarization of $p_x^{\text{edge}}=0.5$ and 
$p_y^{\text{edge}}=0.5$, respectively. Thus, the relation that 
$Q^{\mathrm{corner~}}=p_y^{\mathrm{edge~}}+p_x^{\mathrm{edge~}}-q_{xy}$
is still preserved. 

In summary, we have proposed a model Hamiltonian that demonstrates a novel type of 
quadruple insulator with a nonzero quadrupole moment.
The insulator does not exhibit midgap corner modes in the energy spectrum but does 
have midgap modes in the entanglement spectrum.
Our results indicate that 
the bulk-edge correspondence of quadrupole insulators generally manifests in the 
entanglement spectrum instead of the energy spectrum.  
Our analysis reveals that the breakdown of the relationship between the energy spectrum 
and entanglement spectrum arises because the edge energy gap can close during the
process of flattening of the Hamiltonian while preserving its bulk states.
Importantly, our findings are not restricted to 2Ds as the model can 
be extended to three dimensions (3Ds), identifying octupole insulators devoid of 
midgap modes in the energy spectrum.
Furthermore, it is possible to investigate the semimetallic phase in 3Ds, 
where the presence of hinge arc states is exclusively observed in the entanglement spectrum, 
rather than the energy spectrum.

\begin{acknowledgments}
The work is supported by the National Natural Science Foundation
of China (Grant No. 11974201) and Tsinghua University Dushi Program.
\end{acknowledgments}

\emph{Note added}: During the preparation of this
manuscript, we became aware of a related work~\cite{Yang2023arxiv}.

\begin{widetext}
	\setcounter{equation}{0} \setcounter{figure}{0} \setcounter{table}{0} %
	\renewcommand{\theequation}{S\arabic{equation}} \renewcommand{\thefigure}{S%
		\arabic{figure}} \renewcommand{\bibnumfmt}[1]{[S#1]} 
	\renewcommand{\citenumfont}[1]{S#1}
	
	In the Supplemental Material, we will follow Refs.~\cite{SMTurner2010PRB,SMTHughes2011PRB} to show 
	the relation between the entanglement spectrum in our case and the energy spectrum of the flattened Hamiltonian.  
	
	We divide an $L_x\times L_y$ system into a quarter part $A$ and its complement $\overline{A}$ [see Fig. 3(a) in the main text]
	\cite{SM_Schindler2018SA,SM_Wang2018EL,SM_Hughes2020PRB}.
	The entanglement spectrum refers to the spectrum of the reduced density matrix $\rho_A$ of the subsystem $A$
	by tracing out its complement $\overline{A}$ for the density matrix of the ground state $|\Psi_G\rangle$
	\cite{SMHaldane2008PRL,SMPollmann2010PRB,SMFidkowski2010PRL},
	that is,
	\begin{align} \label{dm}
		\rho_A=\text{Tr}_B(|\Psi_G\rangle\langle\Psi_G|)=\frac{e^{-H_A}}{Z_A}.
	\end{align}
	Here we write $\rho_A$ in terms of a Hamiltonian $H_A$, and $Z_A=\text{Tr}e^{-H_A}$.
	
	For a non-interacting system, the entanglement spectrum is determined by eigenvalues of the single-particle 
	correlation matrix in the region $A$~\cite{SM_Peschel2003}, 
	\begin{align} \label{corre}
		\left[C_A\right]_{\bm{r}_i\alpha,\bm{r}_j\beta}=\langle\hat{c}^\dagger_{\bm{r}_i\alpha}\hat{c}_{\bm{r}_j\beta}\rangle,
	\end{align}
	where $\hat{c}^\dagger_{\bm{r}_i\alpha}$ ($\hat{c}_{\bm{r}_i\alpha}$) is the fermionic creation (annihilation) operator at lattice site 
	$(\bm{r}_i,\alpha)$. Writing in momentum space, we have
	\begin{align} \label{corre_k}
		\left[C_A\right]_{\bm{r}_i\alpha,\bm{r}_j\beta}=\frac{1}{L_x L_y}
		\sum_{\bm{k}}\langle\hat{c}^\dagger_{{\bm{k}}\alpha}\hat{c}_{{\bm{k}}\beta}\rangle e^{-i\bm{k}\cdot(\bm{r}_i-\bm{r}_j)},
	\end{align}
	with $\hat{c}_{{\bm{k}}\alpha}=\frac{1}{\sqrt{L_x L_y}}\sum_{\bm{r}_i}\hat{c}_{\bm{r}_i\alpha}e^{-i\bm{k}\cdot\bm{r}_i}$ 
	denoting the fermionic annihilation operator of component $\alpha$ in momentum space. 
	Diagonalizing the Hamiltonian, we have
	\begin{align} \label{Hksec}
		\hat{H}=\sum_{\bm{k},n}E_n(\bm{k})\hat{f}^\dagger_{{\bm{k}}n} \hat{f}_{{\bm{k}}n},
	\end{align}
	where $\hat{f}^\dagger_{{\bm{k}}n}=\sum_\alpha \hat{c}^\dagger_{\bm{k}\alpha} [u_{\bm{k}}^n]^\alpha $, and 
	$[u_{\bm{k}}^n]^\alpha $ is the $\alpha$th component of the eigenvector of $H(\bm{k})$ with eigenenergy $E_n(\bm{k})$. For $\langle\hat{c}^\dagger_{{\bm{k}}\alpha}\hat{c}_{{\bm{k}}\beta}\rangle$, we have
	\begin{align} \label{corre_ku}
		\langle\hat{c}^\dagger_{{\bm{k}}\alpha}\hat{c}_{{\bm{k}}\beta}\rangle
		&=\langle\Psi_G|\hat{c}^\dagger_{{\bm{k}}\alpha}\hat{c}_{{\bm{k}}\beta}|\Psi_G\rangle \\ \nonumber
		&=\sum_{nn^\prime} [u_{\bm{k}}^{n*}]^\alpha [u_{\bm{k}}^{n^\prime}]^\beta \langle\Psi_G| \hat{f}^\dagger_{{\bm{k}}n} \hat{f}_{{\bm{k}}n^\prime} |\Psi_G\rangle \\ \nonumber
		&=\sum_{n\in \text{occ}.} [u_{\bm{k}}^{n*}]^\alpha [u_{\bm{k}}^{n}]^\beta,
	\end{align}
	which only involves occupied states. Substituting Eq.~(\ref{corre_ku}) into Eq.~(\ref{corre_k}) yields
	\begin{align} \label{cor}
		\left[C_A\right]_{\bm{r}_i\alpha,\bm{r}_j\beta}&=\frac{1}{L_x L_y}\sum_{\bm{k},n\in\text{occ}.} 
		[u_{\bm{k}}^{n*}]^\alpha [u_{\bm{k}}^{n}]^\beta 
		e^{-i\bm{k}\cdot(\bm{r}_i-\bm{r}_j)}.
	\end{align}
	We now consider the projector onto the occupied states $P_{\text{occ}}=\sum_{{\bm k}}\sum_{n\in {\text{occ.} }} |{\bm k} n \rangle \langle {\bm k} n|$.
	Its representation in real space is given by
	\begin{align} \label{proj}
		\left[P_{\text{occ}}\right]_{\bm{r}_i\alpha,\bm{r}_j\beta}&
		=\frac{1}{L_x L_y}\sum_{\bm{k},n\in\text{occ}.} [u_{\bm{k}}^{n}]^\alpha [u_{\bm{k}}^{n*}]^\beta  
		e^{i\bm{k}\cdot(\bm{r}_i-\bm{r}_j)}=\left[C_A\right]_{\bm{r}_i\alpha,\bm{r}_j\beta}^*.
	\end{align}
	Obviously, the correlation matrix $C_A$ is the complex-conjugate matrix of the real-space projector 
	restricted in the subsystem $A$. Since the flattened Hamiltonian is defined by the projector, i.e., 
	$H_{\text{flat}}=\frac{1}{2}-P_{\text{occ}}$,
	then 
	\begin{equation}
		H_{\text{flat}}^t=\frac{1}{2}-C_A,
	\end{equation} 
	where $t$ represents the transpose operation. Thus, the entanglement spectrum has a one-to-one correspondence with the energy 
	spectrum of the flattened Hamiltonian. Specifically, if there is a midgap mode at $\xi=0.5$ for $C_A$, then there exists a zero-energy mode
	for the flattened Hamiltonian, and vice versa.

\end{widetext}

\end{document}